\documentclass[%
 reprint,
amsmath,amssymb,
showkeys,
aps,
]{revtex4-2}

\usepackage{graphicx}
\usepackage{lipsum}
\usepackage{mwe}
\usepackage[bottom]{footmisc}
\usepackage{enumitem}   
\usepackage{dcolumn}
\usepackage{bm}
\usepackage{comment}

\usepackage[]{natbib}
\usepackage{hyperref}

\usepackage{amsmath} 
\usepackage[utf8]{inputenc}
\usepackage{xcolor}

\usepackage[normalem]{ulem}

\begin{document}

\title{{Gravitational-wave asteroseismology with $f$-modes from neutron star binaries at the merger phase}}

\author{Harry Ho-Yin Ng}
\email{hoyin.ng@ligo.org}
\author{Patrick Chi-Kit Cheong}
\email{chi-kit.cheong@ligo.org}
\author{Lap-Ming Lin}
\email{lmlin@phy.cuhk.edu.hk}
\author{Tjonnie Guang Feng Li}
\email{tgfli@cuhk.edu.hk}

\affiliation{Department of Physics, The Chinese University of Hong Kong, Shatin, N.T., Hong Kong}

\date{15 December 2020}

\begin{abstract}
	Gravitational-wave signals from coalescing binary neutron stars can yield important information about the properties of nuclear-matter equation of state from the early part of the signal through tidal effects to the properties and oscillation frequencies of the merger product.
	In this work, we investigate a direct link between the properties of isolated neutron stars and their merger, by comparing the frequency of the fundamental oscillation mode ($f$-mode) of neutron stars with the gravitational-wave frequency associated with the merger of two neutron stars.
	In particular, we calculate the quadrupolar ($l=2$) $f$-mode oscillation ($f_{2f}$) of non-rotating and rotating neutron stars using a nonlinear hydrodynamics code in the conformally-flat approximation and obtain the gravitational-wave frequency associated with the peak amplitude ($f_{\rm max}$) of binary-neutron stars from a set of publicly available simulations.
	We find that $f_{\rm max}$ and $f_{2f}$ differ by about 1\%, on average, across forty-five equal-mass systems with different total mass and equations of state.
	Interestingly, assuming that the gravitational-wave frequency is still approximately equal to twice the orbital frequency $\Omega$ near the merger, the result indicates that the condition for tidal resonance $|m|\Omega = f_{2f}$ is satisfied to high accuracy near the merger, where $m=2$ is the azimuthal quantum number.
	While it has been suggested that the resonance condition could be satisfied near the merger phase, this is the first time that the accuracy of the resonance condition is quantified.
	Moreover, the well established universal relation between $f_{\rm max}$ and the tidal deformability of equal-mass binary systems can now be explained by a similar relation between $f_{2f}$ and the tidal deformability of isolated neutron stars, which has been demonstrated to be associated with the nearly incompressible properties of neutron stars.
	For unequal-mass binaries, $f_{\rm max}$ is generally smaller than the $f$-mode frequencies of the two stars, and the deviation increases as the mass ratio decreases from unity for the limited systems that we have surveyed.
	Therefore, our findings suggest that it is possible to relate the gravitational-wave signal at the merger of a binary neutron star system directly to the fundamental oscillation modes and the mass ratio. This work potentially brings gravitational-wave asteroseismology to the late-inspiral and merger phases of binary neutron stars, filling the gap between the early inspiral and post-merger signals.

\end{abstract}

\maketitle

\section{Introduction}\label{sec:introduction}

\noindent
The observation of gravitational waves (GWs) from binary neutron star (BNS) systems will be an important channel to probe the uncertain properties of the supra-nuclear equation of state (EOS) and may even provide evidence for the existence of deconfined quark matter and phase transitions inside neutron stars (NSs) \cite{yagi2014effective,ranea2018oscillation,flores2019fundamental,barack2019black}.

Studies have shown that different physical effects imprint various signatures in the GW signal during the inspiral and post-merger phases of a BNS system \cite{stergioulas2011gravitational,pratten2020gravitational}. The inspiral phase depends on the EOS through the tidal deformability $\lambda_{l}$ of an NS, which characterizes the multipole moments (order $l$) of the deformation induced by the tidal field of the companion \cite{mora2004post,flanagan2008constraining}.
The tidal deformability is known to be sensitive to the EOS of NS matter \cite{read2013matter,malik2018gw170817}.
The analysis of the first GW observation of a BNS (GW170817) has already put limits on this parameter and provided constraints on various EOSs \cite{abbott2018gw170817,malik2018gw170817,tews2018critical}.

The internal oscillation modes of NSs can be excited by the tidal field during the inspiral. 
The quadrupolar $l = 2$ fundamental fluid mode ($^{2}f$-mode) is expected to have the strongest tidal coupling because it vibrates like standing waves on the surface of the NS and the amplitude extends towards the surface \cite{ho2018gravitational,schmidt2019frequency}.
The coupling between the excited $^{2}f$-modes and tidal fields has been found to lead to dynamical tidal effects, which becomes dominant when the tidal forcing frequency approaches the $^{2}f$-mode frequency \cite{hinderer2016effects}.
This effect results in frequency-dependent tidal deformability (or tidal Love number $k_{l}$) and an extra phase-shift in the emitted GW beyond the static tide limit \cite{hinderer2016effects,steinhoff2016dynamical}.
The inclusion of the dynamical tidal contribution in the analysis of  GW170817 has placed constraints on the $^{2}f$-mode frequency \cite{pratten2020gravitational}.

The $^{2}f$-mode of NSs is also interesting in its own right as its frequency, in general, depends sensitively on the EOS and the spin of the star \cite{andersson1998towards,dimmelmeier2006non}.
For example, the post-merger GW signals due to the $^{2}f$-mode oscillations of the hypermassive NS formed after the merger can also provide us an important channel to study the hot EOS and properties of the merger remnant \cite{stergioulas2011gravitational,bauswein2012measuring, takami2014constraining,rezzolla2016gravitational,Bauswein_2016}.

In this paper, we add a twist to the story by considering, instead of the inspiral and post-merger phases, whether the GW signals {\it at} the merger can be used to probe the $^{2}f$-modes of the two initial NSs.
The merger signal has an obvious advantage in that the GW amplitude is the largest.
While current detectors are not yet sensitive enough in the high frequency range ($\sim 1000$-$2000$ Hz) to detect the merger signals directly, it might not be too far away for current detectors such as Advanced LIGO \cite{aasi2015advanced}, Advanced Virgo \cite{acernese2014advanced}, KAGRA \cite{aso2013interferometer} and future detectors such as Einstein Telescope \cite{punturo2010einstein}.  
Depending on the EOS models, the GW frequency at the merger is in general also somewhat lower than those signals associated with the oscillations of the post-merger hypermassive NS \cite{takami2014constraining,rezzolla2016gravitational}, making it more promising to be detected.

The merger phase is characterized by a maximum in the GW amplitude which in turn signals the end of the inspiral phase.
The frequency at the maximum amplitude $f_{\rm max}$ has been well studied in BNS simulations \cite{rezzolla2016gravitational,takami2015spectral,read2013matter}.
It is also found that there exists a strong correlation between $f_{\rm max}$ and the tidal deformability (expressed by the tidal coupling constant $\kappa_{2}^{T}$).

We consider the numerical data for different BNS systems published in \cite{rezzolla2016gravitational,dietrich2017gravitational,dietrich2017gravitational2}.
Surprisingly, we find that the values of $f_{\rm max}$ for equal-mass binaries agree very well to those of the $^{2}f$-mode frequencies $f_{2f}$ of the two non-rotating and rotating initial NSs that we computed using a hydrodynamics code \cite{cheong2020gmunu}.
In particular, the average relative difference between $f_{\rm max}$ and $f_{2f}$ is about $1\%$ level across the EOS models and mass range ($1.2 - 1.4 M_\odot$) that we have considered. The close relation between $f_{\rm max}$ and $f_{2f}$ leads us to propose that the universal relation (UR) connecting $f_{\rm max}$ and $\kappa_{2}^{T}$ for BNS systems \cite{read2009constraints,bernuzzi2014quasiuniversal,rezzolla2016gravitational}, originates from the UR between the $^{2}f$-mode frequencies and the tidal deformability for isolated NSs as found in \cite{chan2014multipolar}.
For unequal-mass binaries, the correlation between the two frequencies becomes weaker and the largest relative difference in our data set is about $-35.5\%$.
With the limited data set that we have for unequal-mass binaries, we see hints that the difference between the two frequencies increases as the mass ratio decreases from unity.
Our study, therefore, suggests that one can obtain information about the ${2}^f$-mode frequency and mass ratio $q$ of a BNS system using the GW peak frequency $f_{\rm max}$ at the merger phase. Unless otherwise stated, physical quantities are represented in dimensionless units $c = G = M_{\odot} = 1$.

\section{Numerical Setup} 
\label{sec:simulation}
\subsection{Fundamental oscillation mode of rotating neutron stars}

\noindent We follow the approach of Ref.~\cite{dimmelmeier2006non} to determine the frequency of $^{2}f$-mode frequency of an isolated NS by suitably perturbing and following the evolution of the star by a non-linear hydrodynamics code.
We employ an open-sourced code XNS \cite{bucciantini2011general,pili2014axisymmetric}, supplemented with realistic EOSs with piecewise polytropic approximation, to generate axisymmetric uniformly rotating NSs to serve as initial data for our simulations.
We consider 3 sequences of fixed angular velocity $a$ , namely a non-rotating sequence (U0) and two rotating sequences with frequencies $418\,\textrm{Hz}$ (U1) and $673\,\textrm{Hz}$ (U2). 
For each sequence, we consider NSs with different gravitational mass ranging from $M_{g} = 1.20$ to $1.60$ $M_{\odot}$, which are modeled by 6 different realistic cold EOSs: APR4 \cite{akmal1998equation}, SLY \cite{douchin2001unified}, ENG \cite{engvik1995asymmetric}, ALF2 \cite{alford2005hybrid}, GNH3 \cite{glendenning1984neutron} and H4 \cite{glendenning1991reconciliation}. 
These EOSs all satisfy the current observational lower bound on the maximum mass \cite{antoniadis2013massive}.
To save computational time and minimize interpolation errors, we use piecewise polytropic approximations each with 4 pieces instead of tabulated EOSs, which have been shown to reproduce sufficiently accurate representations \cite{read2009constraints}.
In particular, inside the range of $\rho_{i-1} \leq \rho<\rho_{i}$, with $i = 1,2,3,4$,
\begin{align}
	p&=K_{i} \rho^{\mathrm{\Gamma}_{i}},
	 \\
	\epsilon(\rho)&=\left(1+A_{i}\right) \rho+\frac{K_{i}}{\Gamma_{i}-1} \rho^{\Gamma_{i}},
	\\
	A_{i}&=\frac{\epsilon\left(\rho_{i-1}\right)}{\rho_{i-1}}-1-\frac{K_{i}}{\Gamma_{i}-1} \rho_{i-1}^{\Gamma_{i}-1},
\end{align} 
where $\rho$ is the rest-mass density, $\epsilon$ is the energy density, $p$ is the pressure, and the constants $\rho_{i}$ are the cut-off rest-mass density to define the boundaries separating the polytropic models.
The variables $\rho$, $\epsilon$, and $p$ are assumed to be continuous at the boundaries.
The parameters $\Gamma_{i}$ and $K_{i}$ are the adiabatic indices and polytropic constants for the different regions.

We use a new relativistic hydrodynamics code \texttt{Gmunu} \cite{cheong2020gmunu} for the evolution of the initial NS.
\texttt{Gmunu} uses a multigrid method to solve the Einstein equations in the conformally flat condition (CFC) approximation, which has been shown to solve the elliptic-type metric equations resulting in the CFC approximation efficiently.
\texttt{Gmunu} also includes standard high-resolution shock-capturing (HRSC) schemes to solve the hydrodynamics equations, and the Harten-Lax-van Leer-Einfeldt (HLLE) Riemann solver \cite{harten1983upstream} with the TVD reconstruction scheme \cite{kuzmin2004high}. The time update uses a 3rd order Runge-Kutta method. 
While the code is not fully general relativistic, the CFC approximation has been demonstrated to be a good approximation to model various astrophysical problems \cite{saijo2004collapse,dimmelmeier2006non,cordero2009improved,bauswein2012equation,bauswein2014revealing}.

We perform axisymmetric simulations using spherical coordinates $(r,\theta)$ with an equidistantly spaced grid resolution $n_{r} \times n_{\theta} = 320 \times 32$.
In our simulations, the interior of a typical NS model is resolved by about 120 to 200 grid points along the radial direction.
The outside region of the star is covered by an artificial atmosphere with a density $\rho_{atm} \sim 10^{-7} \rho_{c}$, where $\rho_{c}$ is the central density of the star.
The metric is solved at every 50 hydrodynamic steps and we extrapolate the metric in between.
We evolve the initial profiles by \textsl{Gmunu} by adding the following $\theta$-component velocity perturbation to excite the $l=2$, $m_{l} = 0$ axisymmetric $^{2}f$-mode of rotating NSs \cite{dimmelmeier2006non}:
\begin{align}
v_{\theta} = A \sin \left(\pi \frac{r}{r_{\mathrm{s}}(\theta)}\right)\left(3 \cos ^{2} \theta-1\right),
\end{align}
where $r_{s}(\theta)$ is the coordinate radius of the stellar surface and the perturbation amplitude is chosen to be $A = 0.01$.
While this perturbation function is not an exact mode eigenfunction, it is chosen to mimic the angular dependence of the axisymmetric $^{2}f$-mode so that this mode can be excited and dominate other modes during the evolution of a slowly rotating star.
For rapidly rotating stars, other oscillation modes can also be excited or even be stronger than the ${}^2f$-mode that we focus on \cite{dimmelmeier2006non}.
We note that non-axisymmetric modes cannot be studied in our axisymmetric simulations.

The evolution time for each model is around $60-80$ ms ($\sim$ 5-20 spinning cycles) which we have found to be sufficient for extracting the $^{2}f$-mode. The mode is identified and its frequency is obtained by performing a Fourier transform of the non-radial velocity component $v^{\theta}$ at $n_{r} = 80 $, $\theta = \pi/4$ which is ensured to be inside the star (see \cite{chan2014multipolar} for more details).

\subsection{Tidal coupling constant $\kappa_{2}^{T}$}\label{sec:kappaT2}

\noindent The rescaled tidal coupling parameter $\kappa_{2}^{T}$ of a BNS system is defined from the $l=2$ dimensionless tidal Love numbers $k^{A}_{2}$ and $k^{B}_{2}$ of two NSs $A$ and $B$ of the system when they are at infinite separation \cite{damour2010effective}, namely 
\begin{align}
	\kappa_{2}^{T} \equiv 2\left[q\left(\frac{X_{A}}{\mathcal{C}_{A}}\right)^{5} k_{2}^{A}+\frac{1}{q}\left(\frac{X_{B}}{\mathcal{C}_{B}}\right)^{5} k_{2}^{B}\right], \label{eq:kappaT2}
\end{align}
\begin{align}
	q \equiv \frac{M_{B}}{M_{A}} \leq 1,\label{eq:massratio}\\
	\quad X_{A, B} \equiv \frac{M_{A, B}}{M_{A}+M_{B}},
\end{align}
\begin{align}
\mathcal{C}_{A, B} \equiv M_{A, B} / R_{A, B},
\end{align}
where $q$ is the mass ratio, $\mathcal{C}_{A, B}$, $M_{A, B}$ and $R_{A, B}$ are the compactness, gravitational mass, and radius of $A$ and $B$ when they are at infinite separation.
For equal-mass binaries, 
\begin{align}
k_{2}^{A}=k_{2}^{B}=k_{2},
\end{align}
and hence Eq.~(\ref{eq:kappaT2}) becomes
\begin{align}
	\kappa_{2}^{T} \equiv \frac{1}{8} k_{2}\left(\frac{R}{M}\right)^{5}=\frac{3}{16} \lambda_{2}.\label{eq:kappaT2-2}
\end{align}
The quadrupolar tidal deformability $\lambda_2$ is given by
\begin{align}
\lambda_{2} \equiv \frac{2}{3} k_{2} \left(\frac{R} {M}\right)^{5},
\end{align}
where $M = M_{A} = M_{B}$ and $R = R_{A} = R_{B}$.
Unless otherwise noted, we put a bar on top of a variable to denote its average value, such as the average mass $\bar{M}$, for unequal-mass binaries.
We will also use Eq.~(\ref{eq:kappaT2-2}) to define a rescaled tidal coupling constant for an isolated NS.

\section{Results}

\subsection{Comparing the frequencies of the fundamental oscillation modes to the peak GW amplitude} \label{sec:f2f-fmax}
\noindent In this section, we compare the $^{2}f$-mode frequency ($f_{2f}$) of the NSs when they are at infinite separation (see Sec.~\ref{sec:simulation}) to the frequency of the peak GW amplitude from BNS simulations ($f_{\rm max}$) reported in the literature \cite{chan2014multipolar,rezzolla2016gravitational,dietrich2017gravitational,dietrich2017gravitational2} 
In particular, Ref.~\cite{rezzolla2016gravitational} contains 56 BNS merger simulations assuming initially quasi-equilibrium irrotational initial data (52 equal-mass binaries, 4 unequal-mass binaries).
The simulations span a range of average masses $\bar{M} = 1.20 - 1.50 M_{\odot}$ and 6 cold EOSs (i.e. APR4, SLY, ALF2, GNH3, H4, LS220 \cite{lattimer1991generalized}) with an additional thermal contribution.
During the whole inspiral phase, binaries orbit at least 4 orbits or more before the moment of merger (defined as the time when the GW amplitude attained maximum), and till the postmerger phase where a hypermassive NS survived for $ \approx 25$ ms.
The instantaneous frequency of the GW during the inspiral is computed from the phase $\phi$ of the waveform, which is related to the dominant $l=m=2$ mode of the multipolar expansion of the Weyl curvature scalar $\Psi_{4}$ \cite{baiotti2008accurate}.
The values of the GW frequency at the time of merger $f_{\rm max}$ are shown in the Tab.~II of Ref.~\cite{rezzolla2016gravitational}.

Moreover, we use 12 data points of the merger GW frequency $f_{\rm mrg}$ for unequal-mass, non-spinning initial data presented in Tab.~I of Ref.~\cite{dietrich2017gravitational} and Tab.~V of Ref.~\cite{dietrich2017gravitational2} (using R2 data).
The definition of $f_{\rm mrg}$ is the same as $f_{\rm max}$ for unequal-mass binaries. The binaries have average masses $\bar{M} = 1.25 - 1.375 M_{\odot}$, mass ratio $q = 0.571$, $0.667$, $0.8$, and $0.9$ (two for $q = 0.9$, five for $q= 0.8$, six for $q = 0.667$, and two for $q = 0.571$), which are modeled by various EOSs including SLY, ALF2, GNH3, and H4.

\begin{figure}[h]
	\hspace*{-0.5cm}
	\includegraphics[width=9.2cm, height=8.2cm]{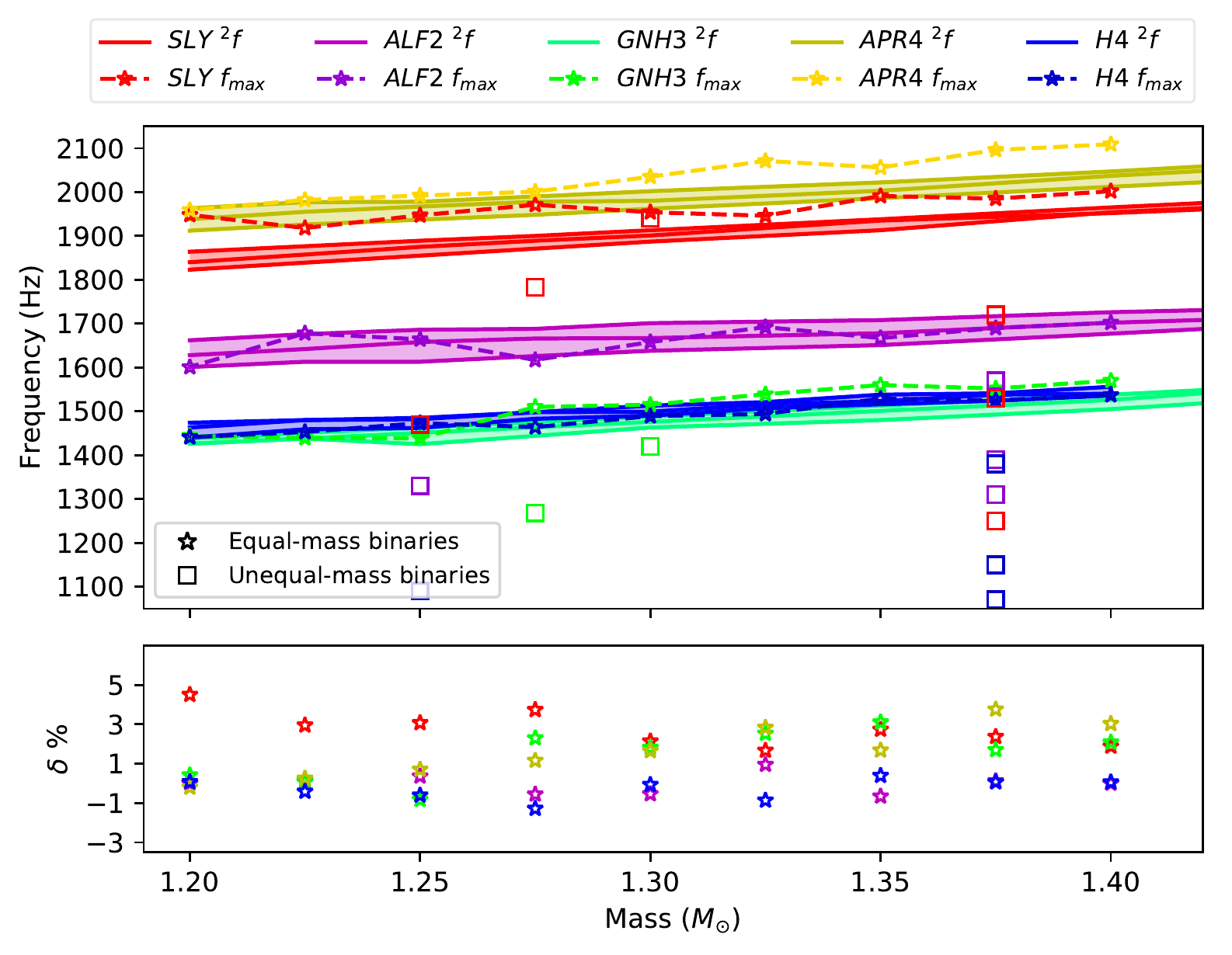}
	\caption{\textit{Upper panel}: 
		The $^{2}f$-mode frequencies $f_{2f}$ (solid line) for three spin sequences (U0, U1, and U2) are plotted against the gravitational mass $M$ of the isolated stars.
		Moreover, the merger gravitational wave peak frequencies $f_{\rm max}$ of equal-mass binaries (star-shaped markers connected by dashed lines) and unequal-mass binaries (square-shaped markers) are plotted against the average mass $\bar{M}$.
		Different colors represent the results from the different EOSs (SLY in red, ALF2 in purple, GNH3 in green APR4 in yellow, and H4 in blue).
	\textit{Lower panel}: Relative difference $\%$ between $f_{\rm max}$ for the equal-mass binaries and the closest value of $f_{2f}$ among the 3 spin sequences with the same EOS and mass.}
	
	\label{fig:2fmodeandfmax}
\end{figure}

Fig.~\ref{fig:2fmodeandfmax} shows $f_{2f}$ against the mass of isolated NS (solid lines) and $f_{\rm max}$ against the average mass of BNS systems (star-shaped markers connected by dashed lines), for various EOSs (colors). 
Note that the EOSs for BNS simulations include an additional thermal contribution to account for the shock heating, while in our simulations of isolated NSs, we use the same set of cold EOSs but without the thermal contribution.

The $^{2}f$-mode data in Fig.~\ref{fig:2fmodeandfmax} consist of three solid lines per EOS (i.e. per color) that correspond to the three spin-frequency sequences (U0, U1, and U2). Therefore, the band for each EOS represents the range of the values of $f_{2f}$ for the range of spin frequency from $0$ to $673$ Hz.
For SLY, APR4, and ALF2 EOSs, we find that $f_{2f}$ increases with the spin frequency so that the non-rotating sequence U0 is given by the bottom solid line in the color bands, while the rotating sequences U1 and U2 are represented by the middle and top solid lines.
On the other hand, the U2 sequence of the GNH3 model has the lowest $f_{2f}$ values and is represented by the bottom line in the green color band.
Finally, the value of $f_{2f}$ for H4 model is somewhat insensitive to the rotation for the range of spin frequency we used so that the band is barely visible.
The strong dependence of $f_{2f}$ on the EOS, spin frequency, and the mass of NS are clearly reflected by the different range and width of the color bands.

Let us begin by focusing on comparison between $f_{\rm max}$ and $f_{2f}$ for equal-mass binaries (star-shaped markers). 
The lower panel in Fig.~\ref{fig:2fmodeandfmax} shows the relative difference $\delta$ between the value of $f_{\rm max}$ and the closest value of $f_{2f}$ of the band for a given EOS and mass $M$.
The average $\delta$ for APR4, SLY, ALF2, GNH3, and H4 are $ +1.65\%$, $ +2.79\%$, $ -0.22\%$, $ +1.47\%$, and $ -0.29\%$ respectively.
We note that, in general, $f_{\rm max}$ has a higher value than $f_{2f}$.
Moreover, we observe that there is a better match between $f_{\rm max}$ and $f_{2f}$ for stiffer EOSs.
The exception is for the softer ALF2, which has the lowest values of $\delta$.
This exception can be explained by the definition of $\delta$, which is the relative difference between $f_{\rm max}$ and the nearest value for $f_{2f}$ among the three rotation profiles for a given mass.
The relatively large spread of the $f_{2f}$ values for ALF2 means that it is more likely to produce a lower value of $\delta$.
For the softest EOSs, SLY and APR4, the difference between $f_{\rm max}$ and $f_{2f}$ are generally larger than those of the stiffer models.
This may be due to the limited spin values that we have used.
Indeed, for soft EOSs, higher spin values typically induce higher $f_{2f}$, so that the relative difference $\delta$ could further decrease by considering faster spinning models. 
Nevertheless, the data in Fig.~\ref{fig:2fmodeandfmax} are sufficient to demonstrate the closeness between $f_{2f}$ and $f_{\rm max}$.

Let us turn to the comparison between $f_{\rm max}$ and $f_{2f}$ for unequal-mass binaries (the relative differences are not shown in the lower panel of Fig.~\ref{fig:2fmodeandfmax} as they, in general, are significantly larger than those of equal-mass binaries).
In particular, the average $\delta$ across data with different mass ratios for SLY, ALF2, GNH3, H4 EOSs are $-16.0\%$ (6 data points), $-15.2\%$ (4 data points), $-7.6\%$ (2 data points) and $ -22.3\%$ (4 data points) respectively. 
We observe that the average $\delta$ becomes more negative as the mass ratio $q$ (less than or equal to unity, as defined in Eq.~(\ref{eq:massratio})) decreases for all EOSs.
In general, the value of $f_{\rm max}$ is lower for unequal-mass binaries than for equal-mass binaries. 
In fact, $f_{\rm max}$ for unequal-mass binaries is even lower than $f_{2f}$ of the lightest NS.
For example, for the case of GNH3 with $q = 0.9$ and $\bar{M} = 1.3 M_{\odot}$ we have $f_{\rm max} = 1420$ Hz and the $^{2}f$-mode frequency of the less (more) massive non-rotating isolated NS with $M = 1.232$ ($M = 1.368$) is $1455$ Hz ($1577$ Hz). 
We observe that the relative difference $\delta$ decreases rapidly as $q \rightarrow 1$.
In particular, the average relative differences are $-6.2 \%$ and $-24.2\%$ for $q \geq 0.8$ and $q < 0.8$ respectively.

The similarity between $f_{2f}$ and $f_{\rm max}$ for equal-mass binaries leads us to conjecture that there may be strong coupling between the ${}^2f$-mode and tidal fields as the tidal forcing frequency near the merger approaches the ${}^2f$-mode frequency.
We will discuss the physical aspects of our results in more detail in Sec.~\ref{sec:conclusions}. However, we first use the similarity between $f_{2f}$ and $f_{\rm max}$ to provide a plausible origin of the $Mf_{\rm max}$ - $\kappa_{2}^{T}$ UR proposed in recent literature \cite{read2013matter,takami2015spectral,rezzolla2016gravitational}.

\subsection{Universal relations}

\begin{figure*} 
	\hspace*{-0.6cm}
		\includegraphics[width=\columnwidth]{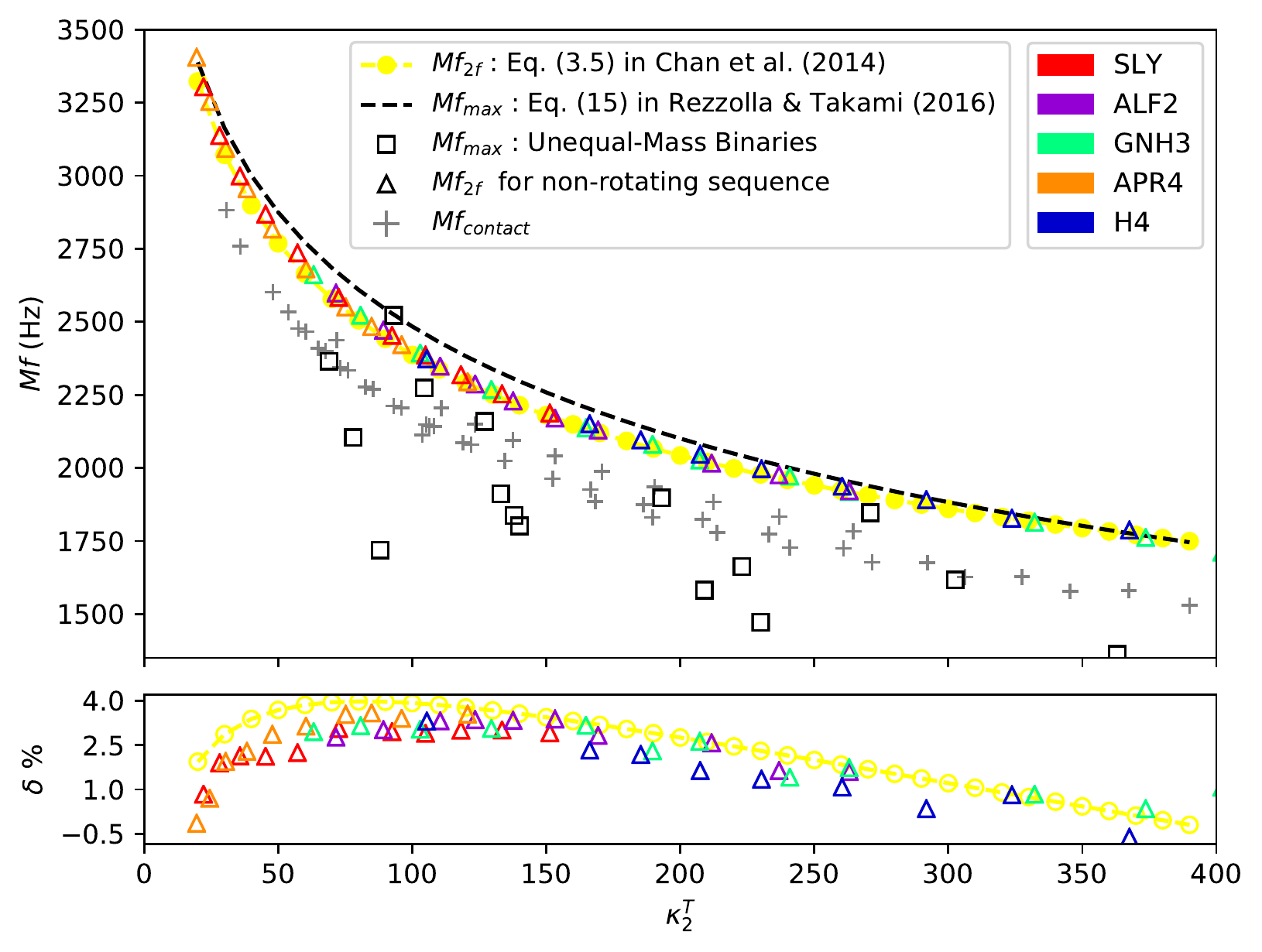}
		\includegraphics[width=\columnwidth]{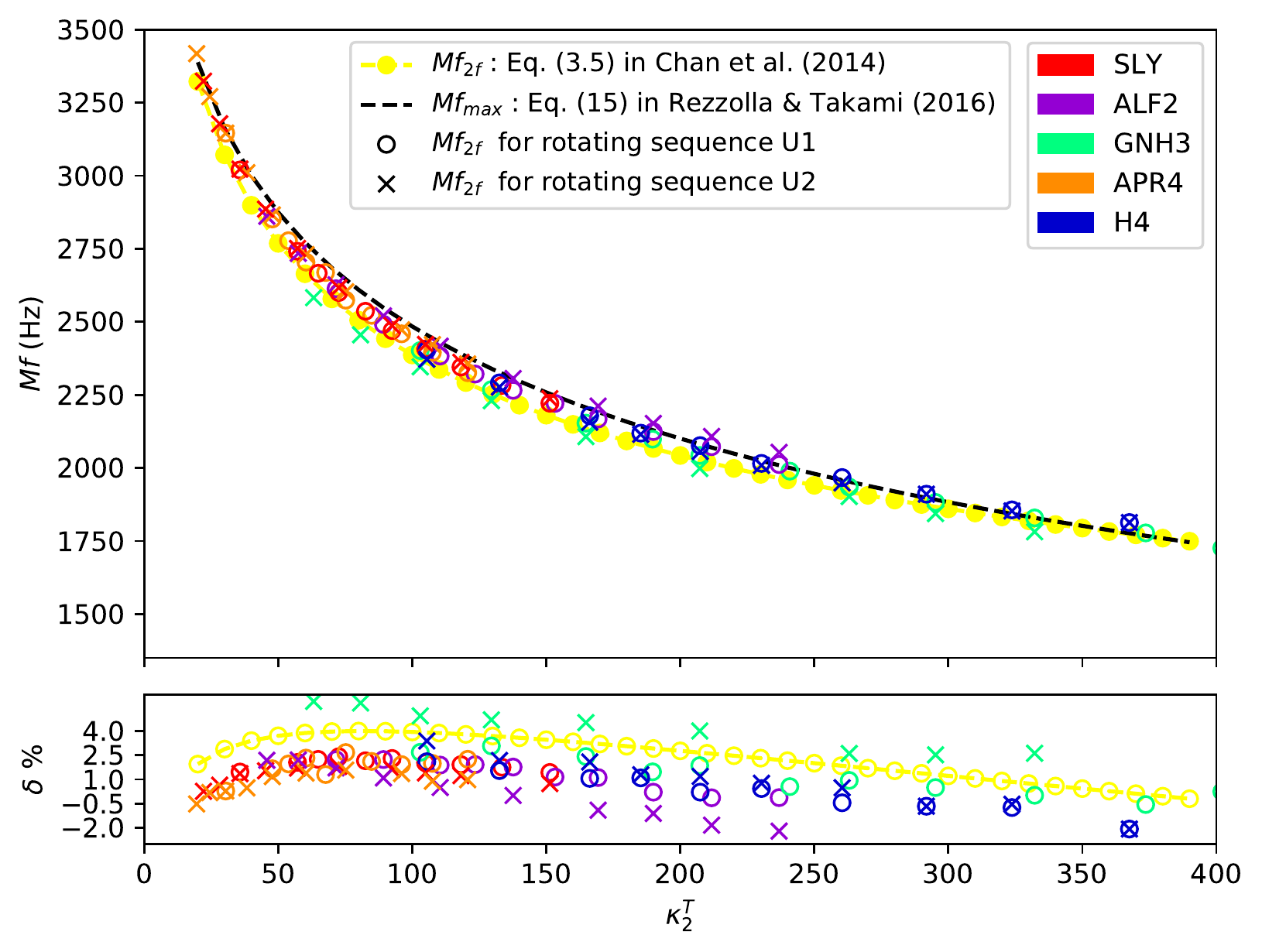}
	\caption{\textit{Upper panel}: 
		The UR (black dashed fit) of the mass-weighted merger GW peak frequencies $M f_{\rm max}$ plotted against the tidal coupling constant $\kappa_{2}^{T}$ obtained from \cite{rezzolla2016gravitational} is shown in both of the plots above. The UR (yellow dotted fit) of the mass-weighted $^{2}f$-mode frequencies of non-rotating NSs plotted against the tidal coupling constant $\kappa_{2}^{T}$ is also shown in the plots above. The mass-weighted $^{2}f$-mode frequencies $M f_{2f}$ for the three spin sequences from U0 to U2 (U0 :$\vartriangle$ in the left figure of Fig$.$ \ref{fig:Mfvslove}, U1:$\circ$ and  U2:$\times$ in the right figure of Fig$.$ \ref{fig:Mfvslove}) are plotted against the tidal coupling constant $\kappa_{2}^{T}$. Different colors represent the results of $M f_{2f}$ from different EOSs. The mass-weighted $\bar{M} f_{\rm max}$ of unequal-mass binaries for different EOSs (black square-shaped markers) obtained from \cite{rezzolla2016gravitational,dietrich2017gravitational,dietrich2017gravitational2} and mass-weighted estimated contact frequency of BNS $M f_{\rm contact}$ for different EOSs (grey plus-shaped markers) defined in \cite{damour2012measurability} are also included. 
		\textit{Lower panel}: Relative differences between the values of mass-weighted $^{2}f$-mode frequency $M f_{2f}$ and the black dashed fit with a given tidal coupling constant $\kappa_{2}^{T}$.}

		\label{fig:Mfvslove}
	
\end{figure*}

Recent studies have empirically established that $f_{\rm max}$ for equal-mass binaries has a strong correlation with the tidal coupling constant $\kappa_{2}^{T}$ through an approximately EOS-insensitive UR \cite{read2013matter, bernuzzi2014quasiuniversal,takami2015spectral,rezzolla2016gravitational}. Interestingly, it has also been found that there exists an UR connecting the $^{2}f$-mode frequency and the tidal deformability of nonrotating cold NSs \cite{chan2014multipolar}.
In this section, by using the finding that $f_{\rm max} \approx f_{2f}$ for equal-mass binaries from Sec.\ref{sec:f2f-fmax}, we provide a plausible origin for the $M f_{\rm max}$ - $\kappa_{2}^{T}$ relation.

There are several slightly different empirical fitting curves for the $M f_{\rm max} - \kappa_{2}^{T}$ relation obtained by different groups \cite{read2013matter,takami2015spectral,rezzolla2016gravitational}. 
In this work, we use the fit given by Eqs. (15) and (17) of Ref.~\cite{rezzolla2016gravitational} and plot it as a black dashed line in Fig.~\ref{fig:Mfvslove}.
For equal-mass binaries, the average deviation between the values of $f_{\rm max}$ and those predicted by the fit is only about $+1.3\%$ \cite{rezzolla2016gravitational}.
Similarly, after replacing $\lambda_2$ by $\kappa_{2}^{T}$ as explained in Sec.~\ref{sec:kappaT2}, the $M f_{2f} - \lambda_2$ fitting curve for nonrotating NSs (cf. Eq. (3.5) of Ref.~\cite{chan2014multipolar}) is plotted as a yellow dotted line in the same figure for comparison.
This $^{2}f$-mode-Love relation is insensitive to EOS models to about $1.0\%$ level \cite{chan2014multipolar}.
It is noted that the black dashed fit is above the yellow fit as the values of $f_{\rm max}$ are slightly higher than those of $f_{2f}$.
The relative differences between the two fits, taking the black dashed fit as a reference, are plotted in the lower panels of Fig.~\ref{fig:Mfvslove}.
While the two URs were discovered independently for different systems (isolated NSs and BNS systems), it is interesting to see that the two fits agree to high accuracy, especially in the small $\kappa_{2}^{T}$ region.
The average deviation between the two fits is about $+2.3\%$ across the range of $\kappa_{2}^{T}$ considered in Fig.~\ref{fig:Mfvslove}.

As a consistency check, we also plot the $f_{2f}$ data for our non-rotating sequence U0 in the left panel of Fig.~\ref{fig:Mfvslove} ($\vartriangle$), which are obtained numerically from hydrodynamics evolutions under the CFC approximation.
We can see that our data agree with the yellow dotted line, which is obtained from a perturbative quasi-normal mode analysis \cite{chan2014multipolar}.
Indeed, the average relative differences ($\delta$) is about $+2.17\%$ (similar to the difference between the two fitting curves) and is insensitive to the EOS.

Moreover, one might argue that the values of $f_{\rm max}$ and $f_{2f}$ could indeed be comparable from the perspective of dimensional analysis.
In Fig.~\ref{fig:Mfvslove}, we also plot the mass-weighted contact frequencies (grey pluses) $M f_{\rm contact}$ against $\kappa_{2}^{T}$ for comparsion.
The contact frequency $f_{\rm contact} = \mathcal{C}^{3 / 2} /(2 \pi M)$ defined in \cite{damour2012measurability}, where $\mathcal{C}$ is the compactness, gives a characteristic GW frequency at the merger.
While the trend of $f_{\rm contact}$ is similar to the black dashed fit, the data are significantly below the fit.
This observation makes it difficult to explain the closeness between the yellow and black fits simply by a dimensional argument. 

The black dashed fit is obtained by fitting 56 BNS simulation data \cite{rezzolla2016gravitational} which includes only 4 unequal-mass systems.
The black dashed fit is thus dominated by equal-mass BNS data.
In Fig.~\ref{fig:Mfvslove}, we also plot the values of ${\bar M} f_{\rm max}$ for unequal-mass binaries presented in \cite{dietrich2017gravitational,dietrich2017gravitational2}.
As already noted in \cite{rezzolla2016gravitational}, the correlation between $f_{\rm max}$ and $\kappa_{2}^{T}$ becomes weaker for unequal-mass binaries.
The deviations (not shown in the lower panel of Fig.~\ref{fig:Mfvslove}) between the data and the black dashed fit are significantly larger for smaller mass ratio $q$.
A lower mass ratio $q$ induces a larger deviation from the black dashed line.
Specifically, the average $\delta$ are $+8.93\%$ and $+25.0\%$ for $q \ge 0.8$ and $q < 0.8$, respectively.
The issue was first stated in Ref.~\cite{rezzolla2016gravitational}, which discussed the ``breaking'' of the UR for small and possibly unrealistic mass ratio.
We explain the ``breaking'' of the $M f_{\rm max}-\kappa_{2}^{T}$ relation by the increase of the deviation between $f_{\rm max}$ and $f_{2f}$ with the decrease of the mass ratio $q$.
With more unequal-mass BNS data available in the future, this could potentially have some observational implications.

Let us now consider the effects of NS rotation.
In the right panel of Fig.~\ref{fig:Mfvslove}, we plot the $^{2}f$-mode data for the spinning sequences U1 ($\circ$) and U2 ($\times$) in addition to the two fits.
The relative differences between the data and the black dashed fit are shown in the lower panel of Fig.~\ref{fig:Mfvslove}.
The average $\delta$ for U1 and U2 are $+1.27\%$ and $+1.17\%$, respectively, which are smaller than the difference ($+2.17\%$) between the two fits.
The results show that the effects of NS spin, in general, can help to shift the $M f_{2f} - \kappa_{2}^{T}$ relation toward the black dashed fit because the values of $f_{2f}$ in general increase with the spin frequency for our chosen EOSs.
One exception is the GNH3 model where the deviations for U1 (green circles) are within $+2.0\%$, while those for U2 (green crosses) can be as larger as $+5.0\%$.
This is due to the fact that the values of $f_{2f}$ of the U2 sequence are smaller than those of the U1 sequence for this EOS as noted above.
For softer EOSs such as SLY and APR4, which are commonly employed in the study of NSs, the U2 data have the least deviations from the black dashed fit.

In summary, we find that the $M f_{\rm max} - \kappa_{2}^{T}$ UR for equal-mass binaries \cite{rezzolla2016gravitational} can be explained by the $M f_{2f} - \kappa_{2}^{T}$ for cold isolated NSs \cite{chan2014multipolar}.
Our conclusion is further strengthened by considering the effects of NS spin on the $^{2}f$-mode frequency.
It is surprising that the two URs can match so well, even though they were originally discovered for quite different systems (isolated vs binary NS systems).
This naturally leads one to ask whether there might be a physical reason for $f_{\rm max} \approx f_{2f}$ which we will try to address in the next section.
Before ending this section, we also remark that the $M f_{2f} - \kappa_{2}^{T}$ UR for isolated NSs can be explained by the nearly incompressible properties of NSs \cite{chan2014multipolar}.
The $M f_{\rm max} - \kappa_{2}^{T}$ relation for BNS systems can now be explained similarly. 

\section{Concluding Remarks}
\label{sec:conclusions}
\noindent In this work, we have investigated the $^{2}f$-mode frequencies of isolated NS with 6 different cold EOSs, 3 sequences of nonrotating and uniformly rotating models in the gravitational-mass range $M = 1.20 - 1.40 M_\odot$.
Interestingly, we found a strong correlation between the $^{2}f$-mode frequencies $f_{2f}$ of isolated NSs and the instantaneous GW peak frequency $f_{\rm max}$ of BNS systems at the merger phase.
We can summarize the main observations as follows:
\begin{enumerate}[label=(\roman*)]
	\item For equal-mass binaries, the ${}^2f$-mode frequencies of the two initial NSs agree very well to the GW peak frequency when the two stars merge.
		The average relative difference between the two frequencies is about $1\%$ across 45 equal-mass systems we have considered.
	\item For unequal-mass binaries, these two frequencies deviate significantly.
		In particular, the difference between $f_{2f}$ and $f_{\rm max}$ increases as the mass ratio $q$ decreases from unity.
		For the limited number of unequal-mass BNS data publicly available to us, we found that the average relative differences between the two frequencies are about $6\%$ and $24\%$ for $q \geq 0.8$ and $q < 0.8$, respectively.
	\item As a direct consequence of $f_{\rm max} \approx f_{2f}$, the $M f_{\rm max} - \kappa_{2}^{T}$ UR for equal-mass binaries \cite{read2013matter,takami2015spectral,rezzolla2016gravitational} can now be explained by the $Mf_{2f} - \kappa_{2}^{T}$ UR discovered for isolated non-rotating NSs, which has been demonstrated to be associated to the nearly incompressible properties of NSs \cite{chan2014multipolar}.
		We also found evidence that rotation can help to improve the agreement between the two URs.
	\item The breaking of the $M f_{\rm max} - \kappa_{2}^{T}$ UR for unequal-mass binaries is a consequence of (ii).
\end{enumerate}

For equal-mass binaries, the observation $f_{\rm max} \approx f_{2f}$ is very interesting and unexpected.
While we cannot rule out the possibility that this result is only a coincidence due to the limited number of BNS data sets that we have surveyed \cite{rezzolla2016gravitational,dietrich2017gravitational,dietrich2017gravitational2}, it is still instructive to ask whether there may be an underlying physical mechanism that is responsible for the observation.
As mentioned in Sec.~\ref{sec:introduction}, the quadrupolar ${}^2f$-mode has the strongest tidal coupling and can be excited by the tidal fields during the inspiral phase of a BNS system.
The coupling between the excited ${}^2f$-mode and tidal fields can lead to complicated dynamical tidal effects \cite{hinderer2016effects,steinhoff2016dynamical} as the tidal forcing frequency approaches the ${}^2f$-mode frequency.
In particular, it is well established that the condition for the oscillation mode to be driven resonantly is given by $|m|\Omega = 2\pi f_{2f}$, where $m$ is the azimuthal quantum number and $\Omega$ is the orbital angular frequency (e.g., \cite{hinderer2016effects}).

For circular binaries, the dominant gravitational-wave frequency $f^{GW}$ during the inspiral is twice the orbital frequency, $f^{GW} = 2 \Omega / 2 \pi$, although higher harmonics with much smaller amplitudes can also exist.
Since the BNS data that we have used in this study are generated from quasi-equilibrium irrotational initial data \cite{rezzolla2016gravitational,dietrich2017gravitational,dietrich2017gravitational2}, it is expected that $f^{GW} = 2 \Omega / 2 \pi$ is satisfied to high accuracy as these binaries maintain quasi-circular orbits during most of the inspiral phase.
However, it is unclear whether this condition still holds in the highly nonlinear dynamical merger phase.
Assuming that $f^{GW}$ does not deviate from $2 \Omega / 2 \pi$ significantly, the observation $f_{\rm max} \approx f_{2f}$ for equal-mass binaries implies that the resonance condition $|m|\Omega /2\pi \approx f_{\rm max} \approx f_{2f}$ is satisfied to high accuracy at the merger phase with $|m|=2$.
While this observation alone does not necessarily mean that the $f$-modes of the stars were driven resonantly in the simulations, the $f$-modes should nevertheless be excited as the tidal driving frequency approaches resonance. 
The resulting dynamical tidal effects on the emitted GW signals have been well studied \cite{hinderer2016effects,steinhoff2016dynamical,andersson2019seismology}.
While it has been suggested that the resonance condition could be satisfied near the merger phase, this is the first time that the accuracy of the resonance condition is quantified.
Furthermore, one open question is whether the $f$-modes could grow to large enough amplitudes to affect the orbital dynamics significantly during the last few orbits.

Motivated by the observation that $f_{\rm max} \approx f_{2f}$, and hence the resonance condition is satisfied to high accuracy at the merger phase across the equal-mass binaries that we have surveyed, we conjecture that the $f$-modes can grow to large amplitudes and lead to strong tidal coupling in the highly nonlinear regime near the merger.
During the last few orbits, the strong tidal coupling draws the orbital energy to the mode oscillations.
When the tidal driving frequency approaches resonance, the energy loss makes the orbit shrink much faster leading to the rapid growth of $\Omega$.
The two NSs merge shortly after the resonance condition is satisfied.
This explains why $f_{\rm max} \approx f_{2f}$, and in particular it is slightly larger than $f_{2f}$, to high accuracy at the moment of merger for all the equal-mass binaries that we have surveyed.

In contrast to the situations of spinning \cite{wynn1999resonant,Ma_2020} and eccentric binaries 
\cite{Yang_2019,Vick_2019}, where the tidal $f$-mode resonance has been well studied, it is generally not expected that the $f$-modes in non-spinning circular binaries can be resonantly excited due to their high frequencies \cite{lai1994resonant,wynn1999resonant}. It should also be noted that the resonances of other low-frequency modes, such as the g-modes and r-modes, have also be studied \cite{shibata1994effects,lai1994resonant,wynn1999resonant,flanagan2007gravitomagnetic}. Our conjecture about the properties of BNS systems in the highly nonlinear merger phase does not necessarily contradict the general expectation based on the linear mode analysis in Newtonian and post-Newtonian approximations. Most importantly, our conjecture should be falsifiable by existing BNS simulation data. By following the fluid motions of the stars in a BNS simulation, one should in principle be able to monitor the growth of the $f$-mode oscillations (if exist). We notice that such an investigation of the $f$-mode oscillations has been done for eccentric BNS simulations \cite{gold2012eccentric}. Being able to see rapid growth of the $f$-mode oscillations in the last few orbits near the merger could provide support to our conjecture. On the other hand, the conjecture can be disproved directly if there is no evidence of the $f$-mode oscillations in the equal-mass BNS simulations that we have surveyed. \\

For unequal-mass binary cases, we have observed that the values of $f_{\rm max}$ and $f_{2f}$ differ from each other significantly for the limited number of data that we have studied, and the difference increases as the mass ratio decreases. 
In addition, $f_{\rm max}$ is smaller than both values of $f_{2f}$ for the two stars. A possible reason is that the exact moment of merger could not be simply defined as the time of GW signal at maximum amplitude due to the asymmetry of the system, because the smaller mass companion may be tidally disrupted by the more massive companion before the merger.
With the disruption of the smaller mass companion, the tidal couplings and excitations of the $f$-modes become less significant comparing to equal-mass binaries. BNS systems with smaller mass ratio may also result in an earlier disruption of the smaller mass companion, and hence lead to a larger difference between $f_{\rm max}$ and $f_{2f}$. Nevertheless, more simulation results are needed to provide a detailed investigation of this issue. \\

Advanced LIGO, Advanced Virgo, KAGRA and third-generation GW detectors such as Einstein Telescope will be able to cover the sensitivity band with $1000 - 3000$ Hz, making it possible to detect $f_{\rm max}$ which has the largest amplitude, from the GW signals of BNS systems in the future.
Our observation that $f_{\rm max} \approx f_{2f}$ for equal-mass binaries can be used to yield important information about BNS systems and provide constraints on EOSs.
For instance, assuming that the mass ratio ($q \geq 0.9$) of a nearly equal-mass binary can be inferred accurately during the inspiral phase and $f_{\rm max}$ can also be detected at the merge phase of the same system, then the $f$-mode frequencies of the stars can be inferred directly.
As the $f$-mode frequency depends sensitively on the EOS, this can help to put constraints on the EOS models.
Using the $Mf_{2f} - \kappa_{2}^{T}$ UR of \cite{chan2014multipolar}, one can also infer the tidal deformability from the merger signal and provide a consistency check for the same physical quantity that may be inferred from the GW signals observed in the inspiral phase.
On the other hand, one can also turn the argument around and use the tidal deformability inferred in the inspiral phase to constrain the $f$-mode frequency \cite{pratten2020gravitational} and compare with $f_{\rm max}$.
The difference between the two frequencies could provide a way to constrain the mass ratio from the merger signals. However, more unequal-mass BNS simulation data are needed to study how the frequency difference $\delta f$ varies with the mass ratio and EOS models. 

While the GW signals in the post-merger phase are strongly associated with the normal mode oscillations of the hypermassive star formed in the merger \cite{bauswein2012measuring,takami2014constraining,Bauswein_2016}, it is surprising that the GW frequency at the moment of merger can be correlated so well to the $f$-mode frequencies of the initial NSs for nearly equal-mass binaries. Our work potentially brings GW asteroseismology to the late-inspiral and merger phases of BNS systems, filling the gap between the early inspiral and post-merger signals.

\section*{Acknowledgements}
\thispagestyle{empty}
The authors thank Alan Tsz-Lok Lam and Chun-Lung Chan for useful discussions and comments on the manuscript.
This work was partially supported by grants from the Research Grants Council of the Hong Kong (Project No. CUHK 24304317 and CUHK 14306419), the Croucher Innovation Award from the Croucher Foundation Hong Kong, and by the Direct Grant for Research from the Research Committee of the Chinese University of Hong Kong.

	\bibliography{ppp_new} 
\end{document}